\documentclass[prd,longbibliography,nofootinbib,twocolumn,preprintnumbers]{revtex4-1}
\usepackage{amsmath,amssymb,amsthm}
\usepackage{graphicx}
\usepackage[usenames, dvipsnames]{xcolor}
\usepackage{slashed}
\usepackage{subfig}
\usepackage{tabularx}
\usepackage{empheq}
\usepackage{mathrsfs}
\usepackage{comment}
\usepackage{bbold}
\usepackage[shortlabels]{enumitem}

\usepackage{url}
\usepackage{color}

\definecolor{light-gray}{gray}{0.9}
\definecolor{medium-gray}{gray}{0.7}
\definecolor{darkblue}{rgb}{0.0,0.0,0.6}
\definecolor{red}{rgb}{0.9, 0,0}
\definecolor{navy}{rgb}{0.05, 0.05,0.8}
\definecolor{linkcolor}{rgb}{0.0, 0.28, 0.67}
\definecolor{paleblue}{rgb}{0.69, 0.93, 0.93}  

\usepackage[colorlinks]{hyperref}
\hypersetup{
     colorlinks   = true,
     citecolor    = linkcolor,
     linkcolor    = linkcolor,
     urlcolor    = linkcolor
}
\newcommand{\be}{\begin{equation}}
\newcommand{\ee}{\end{equation}}

\newcommand{\x}{\chi}

\newcommand{\Ap}{A^\prime}
\newcommand{\mAp}{m_{A^\prime}}
\newcommand{\Lag}{\mathscr{L}}
\newcommand{\eps}{\epsilon}

\newcommand{\p}{\prime}

\newcommand{\grad}{\nabla}

\newcommand{\Eq}[1]{Eq.~\ref{eq:#1}}

\newcommand{\Eqs}[2]{Eqs.~\ref{eq:#1} and \ref{eq:#2}} 
\newcommand{\Sec}[1]{Sec.~\ref{sec:#1}}
\newcommand{\Secs}[2]{Secs.~\ref{sec:#1} and \ref{sec:#2}} 

\newcommand{\Fig}[1]{Fig.~\ref{fig:#1}}

\newcommand{\bebox}{\begin{empheq}[box=\fcolorbox{light-gray}{light-gray}]{align}}
\newcommand{\eebox}{\end{empheq}}

\newcommand{\rhodm}{\rho_{_\text{DM}}}
\newcommand{\DM}{{_\text{DM}}}

\newcommand{\gvec}{\boldsymbol{g}}

\newcommand{\eV}{\text{eV}}
\newcommand{\keV}{\text{keV}}
\newcommand{\MeV}{\text{MeV}}
\newcommand{\GeV}{\text{GeV}}
\newcommand{\TeV}{\text{TeV}}

\newcommand{\cm}{\text{cm}}

\newcommand{\km}{\text{km}}

\newcommand{\Hz}{\text{Hz}}

\begin{document}

\preprint{FERMILAB-PUB-24-0353-T}

\title{High-Energy Neutrinos From Millicharged Dark Matter Annihilation in the Sun}

\author{Asher Berlin$^{a,b}$}
\email{aberlin@fnal.gov}
\author{Dan Hooper$^{a,c,d}$}
\email{dhooper@uchicago.edu}
\affiliation{$^a$Theory Division, Fermi National Accelerator Laboratory}
\affiliation{$^b$Superconducting Quantum Materials and Systems Center (SQMS),
Fermi National Accelerator Laboratory}
\affiliation{$^c$University of Chicago, Department of Astronomy and Astrophysics}
\affiliation{$^d$University of Chicago, Kavli Institute for Cosmological Physics}


\begin{abstract}
Millicharged dark matter particles can be efficiently captured by the Sun, where they annihilate into tau leptons, leading to the production of high-energy neutrinos. In contrast to the Earth, the high temperature of the Sun suppresses the fraction of millicharged particles that are bound to nuclei, allowing for potentially high annihilation rates. We recast existing constraints from the IceCube Neutrino Observatory and use this information to place new limits on the fraction of the dark matter that is millicharged. This analysis excludes previously unexplored parameter space for masses of $m_\chi \sim (5-100) \ \text{GeV}$, charges of $q_\chi \sim 10^{-3}-10^{-2}$, and fractional abundances as small as $f_{_\text{DM}} \sim 10^{-5}$. 
\end{abstract}

\maketitle

\section{Introduction}
\label{sec:intro}

Although all known particles carry values of electric charge that are of order unity (or are electrically neutral), 
models featuring elementary particles with much smaller quantities of charge have received a significant amount of interest in recent years~\cite{Haas:2014dda,Izaguirre:2015eya,Magill:2018tbb,Kelly:2018brz,ArgoNeuT:2019ckq,Harnik:2019zee,ArgoNeuT:2019ckq,Oscura:2023qch,Plestid:2020kdm,Bloch:2020uzh,Foroughi-Abari:2020qar,Harnik:2020ugb,Marocco:2020dqu,Berlin:2023gvx,Chen:2022abz,Budker:2021quh,ArguellesDelgado:2021lek,Afek:2020lek}. Such millicharged particles can arise, for example, if an effective charge is generated through the kinetic mixing of the photon with a new light dark photon, $\Lag \supset (\eps / 2) F^{\mu \nu} F'_{\mu \nu}$, where $F^\p_{\mu \nu}$ is the dark photon field strength and $\eps$ is a small dimensionless parameter. This mixing causes particles, $\x^\pm$, charged under this new $U(1)^\p$, to acquire an effective electric charge, $q_\x = \eps \, e^\p/e$, where $e^\p$ is the $U(1)^\p$ gauge coupling and $e$ is the standard electromagnetic coupling~\cite{Holdom:1985ag}. 

Within the context of effective field theory, any value of $\eps \ll 1$ is technically natural. If the Standard Model is embedded within a Grand Unified Theory, 
however, this mixing is generated only through loops of particles that carry both hypercharge and $U(1)^\p$ charge. At the one-loop level, the expected size of this mixing is given by
\be
\eps \sim \frac{e^\p e}{16 \pi^2} \ln\bigg(\frac{M'}{M}\bigg)
\simeq 6 \times 10^{-4} \times \bigg(\frac{e^\p}{e}\bigg) \, \ln\bigg(\frac{M'}{M}\bigg)
\, ,
\ee
where $M'/M$ is the ratio of the masses in the loop~\cite{Holdom:1985ag,Cohen:2010kn,Baumgart:2009tn}. From this perspective, particles with $q_\x \sim 10^{-3}$ are particularly well motivated.

In this work, we consider the possibility that a fraction of the cosmological dark matter density consists of millicharged particles. Although observations of the cosmic microwave background and measurements of the primordial element abundances have been used to place strong constraints on charged dark matter, these observations remain consistent with the possibility that a small fraction, $f_\DM \lesssim 10^{-2}$, of the dark matter could carry significant electric charge~\cite{Dubovsky:2003yn, dePutter:2018xte, Kovetz:2018zan, Buen-Abad:2021mvc,Stebbins:2019xjr}. In particular,
a millicharged dark matter subcomponent could arise as a thermal relic of the early universe. Assuming standard thermal freeze-out, the requirement of perturbative unitarity~\cite{Griest:1989wd} can be used to place the following lower limit on the fractional abundance of this component:
\be
f_\DM \gtrsim 10^{-10} \times (m_\x / \GeV)^2
~,
\ee
where $m_\x$ is the mass of the millicharged particle, $\x^\pm$.

Even for very small fractional abundances, underground searches for dark matter scattering have been used to place extremely strong constraints on millicharged subcomponents, testing particles with charges as small as $q_\x \sim 10^{-10} / f_\DM^{1/2}$~\cite{Emken:2019tni}. These experiments are almost completely insensitive to particles with sufficiently large values of $q_\x$, however, since such particles scatter many times in the terrestrial overburden, shedding their kinetic energy and thermalizing down to room temperature, $300 \ \text{K} \simeq 26 \ \text{meV}$, well below the threshold of typical large-scale experiments. As a result, millicharged particles with $q_\x \gtrsim 10^{-4} \times (m_\x / \GeV)^{1/2}$ reside within a blind spot of existing direct detection efforts~\cite{Emken:2019tni}.

The efficient scattering of millicharged particles with astrophysical objects within our Solar System has a drastic effect on their local distribution and density. For $m_\x \gg 1 \ \GeV$, the characteristic thermal velocity of these particles will be well below the gravitational escape velocity of the Sun or Earth. As a result, such particles remain gravitationally bound to these objects after scattering, accumulating as a dense hydrostatic gas over the age of the Solar System. 

The possible accumulation of dark matter particles within stellar bodies has been studied extensively, dating back to the 1980s~\cite{1985ApJ...299.1001K,Press:1985ug,Griest:1986yu,Gould:1987ir,Gould:1987ju,Gould:1989hm}. It has also long been appreciated that such particles could give rise to interesting signatures, arising from their annihilation into Standard Model states. Examples include the prompt annihilation into neutrinos~\cite{Silk:1985ax,Srednicki:1986vj,Kamionkowski:1991nj,Barger:2001ur,Flacke:2009eu} or long-lived mediators~\cite{Schuster:2009au,Batell:2009zp} in the Sun, annihilations in the local volume of terrestrial neutrino detectors~\cite{Pospelov:2020ktu,McKeen:2023ztq}, and annihilations in Earth's core~\cite{Mack:2007xj,Pospelov:2023mlz}.  

Signals arising from the annihilation of millicharged particles gravitationally bound to Earth was studied in Ref.~\cite{Pospelov:2020ktu}. However, in this case, an additional complication arises from the fact that $\x^-$ can bind with atomic nuclei, $N$. In turn, the resulting Coulomb barrier between $(\x^- N)$ and $\x^+$ drastically suppresses the local annihilation rate of millicharged particles. For $q_\x \gg 10^{-5}$, the binding energy of $(\x^- N)$ greatly exceeds the temperature of Earth's core, exponentially suppressing the rate of terrestrial-based annihilation signals. 

In this paper, we instead focus on the capture and annihilation of millicharged particles in the Sun, leading to the production of high-energy neutrinos through $\x^+ \x^- \to \tau^+ \tau^-$ followed by the decay of the tau leptons into neutrinos. In contrast to millicharged particles in the Earth, the high temperature of the solar interior suppresses the fraction of these particles that will be bound to nuclei, making this an ideal environment to produce detectable fluxes of annihilation products. We recast the results of existing searches for high-energy neutrinos from the Sun, as recently performed by the IceCube Collaboration~\cite{IceCube:2016dgk,IceCube:2021xzo}. Using this data, we place new limits on millicharged dark matter, excluding previously unexplored parameter space corresponding to $m_\x \sim (5-100) \ \GeV$ and $q_\x \sim 10^{-3} - 10^{-2}$ for fractional abundances as small as $f_\DM \sim 10^{-5}$. 

The remainder of this paper is organized as follows. In \Secs{capture}{annihilation}, we discuss the dynamics associated with the capture and annihilation of millicharged dark matter in the Sun. \Sec{bound}~addresses modifications that can arise if $\x^-$ can efficiently form bound states with solar nuclei. This formalism is applied in \Sec{icecube} to derive constraints, using existing data from the high-energy neutrino telescope, IceCube. We then compare these limits to other bounds on millicharged dark matter. In \Sec{conclusion}, we summarize our results and discuss directions for future work.

\section{Capture and Annihilation of Dark Matter in the Sun}
\label{sec:capture}

The total number millicharged particles that are gravitationally bound to the Sun, $N_\x$, evolves in time according to
\be
\label{eq:dNdt}
\frac{d N_\x}{d t} \simeq \Gamma_\text{cap} - N_\x^2 \, K_\text{ann}
~,
\ee
where the first and second terms on the right-hand side correspond to the capture and annihilation rates, respectively. Here, we have neglected the effects of millicharged particle evaporation, since for $m_\x \gtrsim 1 \ \GeV$ their thermal velocities are smaller than the gravitational escape velocity near the Sun's surface, $v_\text{esc} = (2G M_{\odot}/R_{\odot})^{1/2} \simeq 2 \times 10^{-3}$, where $M_{\odot}$ and $R_{\odot}$ are the solar mass and radius, respectively.  

Within the parameter space that we will consider in this work, nearly every millicharged particle that passes through the Sun will scatter, become gravitationally bound, and reach kinetic equilibrium~\cite{Pospelov:2020ktu,Berlin:2023zpn}. To determine the capture rate, we therefore simply calculate the rate at which such particles enter the Sun. The gravitational pull of the Sun significantly increases the flux of dark matter particles that reach its surface. The distance of closest approach to the Sun for a dark matter particle entering the Solar System with an impact parameter, $b$, and a velocity, $v_\DM$, is given by
\be
R_\text{min} = \frac{G \, M_{\odot}}{v^2_\DM} \, \bigg[\bigg(1+\frac{b^2 \, v^4_\DM}{G^2 M^2_{\odot}}\bigg)^{1/2} -1 \bigg]
\, .
\ee
Setting $R_\text{min} \leq R_{\odot}$, we find that such particles will pass through the Sun for all impact parameters less than
\be
b \leq b_\text{max} = R_{\odot} \, \bigg(1+\frac{v_\text{esc}^2}{v^2_\DM}\bigg)^{1/2}
\, .
\ee
From this, it follows that the effective cross-sectional area for capture in the Sun will be given by
\begin{align}
\label{eq:Acap}
A_\text{cap} = \pi \, b_\text{max}^2 =  \pi  R^2_{\odot} \bigg(1+\frac{v_\text{esc}^2}{v^2_\DM}\bigg)
\, .
\end{align}
This leads to a capture rate of
\begin{align}
\label{eq:GammaCap1}
\Gamma_\text{cap} &\simeq (f_\DM \, \rhodm / m_\x) \, v_\DM \, A_\text{cap}
\, ,
\end{align}
where $\rhodm \simeq 0.4 \ \GeV / \cm^3$ is the local dark matter density. To account for the spread in dark matter velocities, we average the quantity $v_\DM A_\text{cap}$ in \Eq{GammaCap1} over the velocity distribution of the Standard Halo Model, consisting of a boosted-Maxwellian (with $v_0 = 220  \ \km / \text{s}$ and $v_{\odot}=240 \ \km / \text{s}$) truncated at the Galactic escape velocity, $v_\text{esc}^{\rm Gal} = 550 \ \km / \text{s}$. We find that this can be approximated by simply using \Eqs{Acap}{GammaCap1} with a representative value of $v_\DM \simeq 9 \times 10^{-4}$.

\begin{figure*}[t]
\includegraphics[width=0.49 \textwidth]{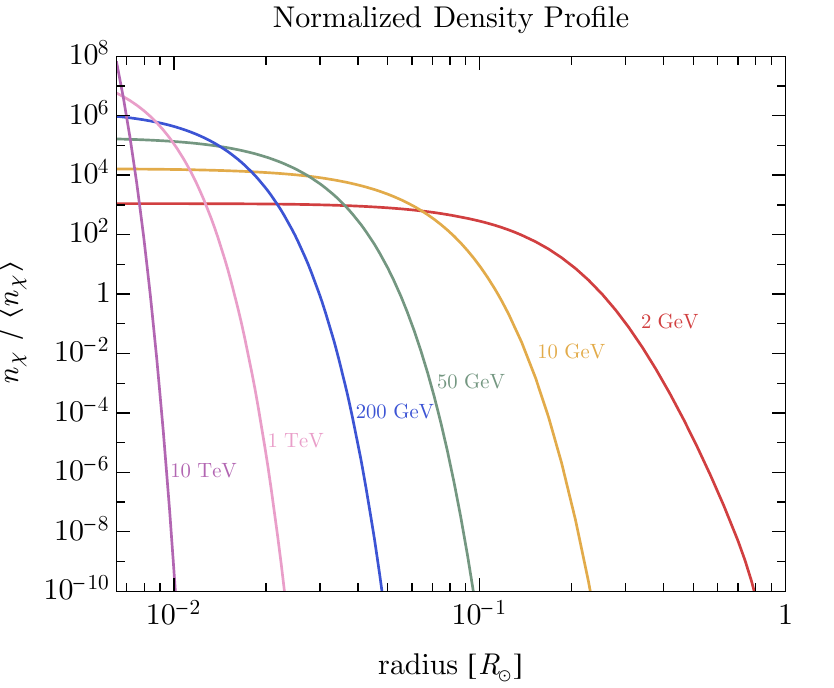}
\includegraphics[width=0.49 \textwidth]{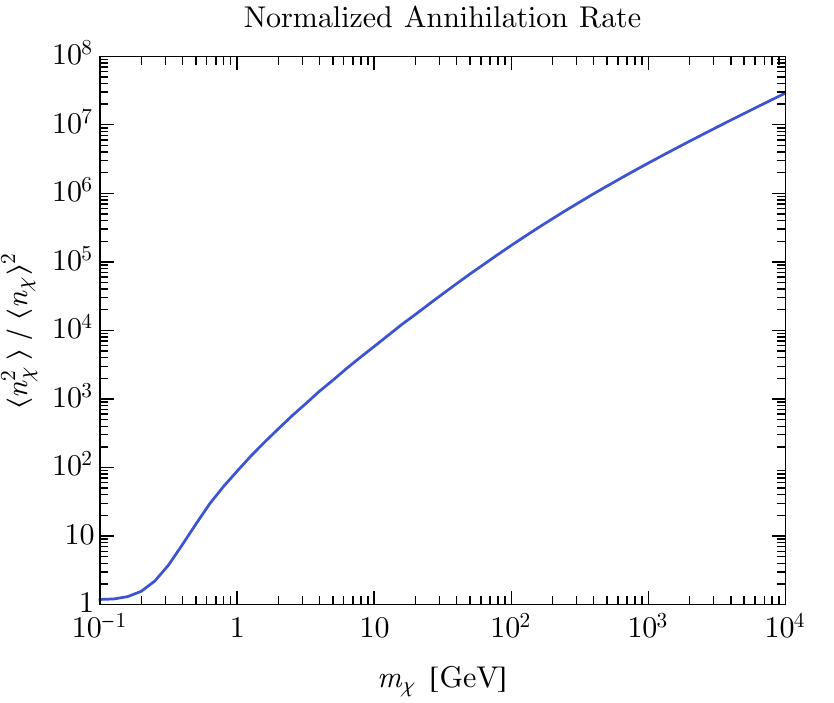}
\caption{\textbf{Left panel}: The density profile of millicharged particles gravitationally bound to the Sun, normalized to the volume-averaged density, $\langle n_\x\rangle$, for several values of the particle's mass. For $m_\x \lesssim 1 \ \GeV$, such particles are approximately uniformly distributed throughout the solar interior. For larger masses, the density profile is much more highly concentrated near the solar core. \textbf{Right panel}: The normalized annihilation rate, $\langle n_\x^2 \rangle / \langle n_\x \rangle^2$, as found in \Eq{Iann}.
For $m_\x \ll 1 \ \GeV$, the uniform density profile of millicharged particles implies that $\langle n_\x^2 \rangle / \langle n_\x \rangle^2 \simeq 1$. For larger masses, the concentration of millicharged particles near the solar core enhances the likelihood of annihilation.}
\label{fig:profile}
\end{figure*}

For a charge-symmetric population of millicharged particles, the quantity $K_\text{ann}$ appearing in \Eq{dNdt} is given by
\begin{align}
\label{eq:Iann}
K_\text{ann} &= \frac{1}{2 N_\x^2} \, \int_{V_\odot} \hspace{-0.1 cm} n_\x^2 \, \langle \sigma v \rangle \, dV
 \simeq \frac{\langle \sigma v \rangle}{2 \, V_\odot} \times \frac{\langle n_\x^2 \rangle}{\langle n_\x \rangle^2}
\, ,
\end{align}
where the integral in the first equality is performed over the volume of the Sun, $V_\odot$, $n_\x \equiv n_{\x^+}+n_{\x^-}$ is the number density of millicharged particles,  and $\langle \sigma v \rangle$ is the thermally-averaged $\x^+ \x^-$ annihilation cross section. In the second equality of this expression, we have approximated $\langle \sigma v \rangle$ as independent of temperature, and the brackets for $\langle n_\x \rangle$ and $\langle n_\x^2 \rangle$ denote an average over the solar volume.  

The propagation of charged particles in the Solar System can be impacted by the turbulent solar wind and its embedded magnetic field, the effects of which are collectively known as solar modulation~\cite{Gleeson:1968zza,Strauss,Maccione:2012cu,Cholis:2015gna,Cholis:2020tpi}. Solar modulation significantly suppresses the intensity of the cosmic ray spectrum at the location of Earth for kinetic energies below $\sim E_\text{kin} \lesssim |Q|\Phi$, where $Q$ is the charge of the cosmic ray and $\Phi \sim (0.2-1) \ \GeV$~\cite{Potgieter:2013mcc,Cholis:2015gna}. 

Analogously, it is possible that solar modulation could suppress the flux of millicharged dark matter particles that reaches the inner Solar System. We expect these effects to be significant if $m_\x \, (v_{\DM}^2+v^2_\text{esc}) / 2 \lesssim q_\x \, \Phi$, or equivalently,
\be
m_\x \lesssim 100  \ \GeV \times \bigg(\frac{q_\x}{10^{-3}}\bigg) \bigg(\frac{\Phi}{0.2 \ \GeV}\bigg)
\, .
\ee
This condition will be satisfied in much of the parameter space that we are interested in, potentially reducing the sensitivity of any solar or terrestrial-based searches for such particles. 

In the above discussion of solar modulation, we have explicitly assumed that the millicharged particles couple to electromagnetic fields in the same way as ordinary charged particles. If these interactions are instead mediated by a kinetically-mixed dark photon, $\Ap$, their range will be limited to $\sim \mAp^{-1}$, where $\mAp$ is the mass of the dark photon. In this work, we will focus on charges, $q_\x = \eps \, e^\p / e \gtrsim 10^{-4}$, corresponding to kinetic mixing parameters of $\eps \gtrsim 10^{-5} / e^\p$. For light dark photons ($\mAp \lesssim \MeV$), such values of $\eps$ are highly constrained unless $\mAp \lesssim 10^{-14} \ \eV \sim (10^{4} \ \km)^{-1}$~\cite{Mirizzi:2009iz,Mirizzi:2009nq,Kunze:2015noa,Caputo:2020bdy,Caputo:2020rnx,Garcia:2020qrp}, corresponding to distances larger than an Earth radius. Hence, within the viable region of parameter space, $\x^\pm$ efficiently couples to terrestrial magnetic fields.
On the other hand, since the coherence length of the solar magnetic field is of the same order as the solar radius~\cite{Longair_2011}, the effects of solar modulation will be suppressed if $\mAp \gg R_\odot^{-1} \sim 10^{-16} \ \eV$. Throughout the remainder of this work, we will implicitly assume that this is the case, allowing us to provide a meaningful comparison between the signals discussed here and other local searches for millicharged dark matter (all of which are subject to the effects of solar modulation). Alternatively, solar modulation may be implicitly incorporated into our analysis by rescaling the incoming
millicharged particle flux, and thereby the definition of $f_\DM$.

To evaluate the annihilation rate in \Eq{Iann}, we will need the radial density profile of millicharged particles in the Sun, $n_\x$. From the condition of hydrostatic equilibrium, the Euler equation implies that
\be
\label{eq:hydrostatic0}
m_\x \, n_\x \, \gvec_\odot = \grad P_\x
~,
\ee
where $\gvec_\odot$ is the solar gravitational field and $P_\x$ is the pressure of the millicharged particles. Treating these particles as an ideal gas, the pressure is given by $P_\x \simeq T \, n_\x$, where $T$ is the temperature of the Sun. In the spherically-symmetric limit, \Eq{hydrostatic0} can be written as
\be
\label{eq:hydrostatic1}
\frac{1}{n_\x(r)} \, \frac{d n_\x (r)}{dr} + \frac{1}{T(r)} \, \frac{d T(r)}{dr} + \frac{m_\x \, g_\odot (r)}{T(r)} \simeq 0
~.
\ee

For a given value of $m_\x$, we solve \Eq{hydrostatic1} numerically, incorporating the solar density and temperature profiles from the standard solar model~\cite{Bahcall:2004fg}. These solutions are presented in \Fig{profile}. In the left panel, we show the normalized density profile, $n_\x (r) / \langle n_\x \rangle$, for several choices of $m_\x$. For $m_\x \lesssim 1 \ \GeV$, gravitationally bound millicharged particles are distributed approximately uniformly throughout the solar interior. For heavier particles, larger pressure and density gradients are required to offset the gravitational attraction, leading to a profile that is much more highly concentrated near the solar core. This behavior is also reflected in the right panel of \Fig{profile}, in which we show the normalized annihilation rate, $\langle n_\x^2 \rangle / \langle n_\x \rangle^2$, as appears in the last factor of \Eq{Iann}. For $m_\x \ll 1 \ \GeV$, the uniform density profile of millicharged particles implies that $\langle n_\x^2 \rangle / \langle n_\x \rangle^2 \simeq 1$. For larger masses, the concentration of millicharged particles near the solar core enhances the likelihood of annihilation, such that $\langle n_\x^2 \rangle / \langle n_\x \rangle^2 \gg 1$.

We are now in a position to calculate the capture and annihilation rates of millicharged particles in the Sun. Solving \Eq{dNdt} with the initial condition $N_\x  = 0$ at $t = 0$, the value of $N_\x$ today is given by
\begin{align}
\label{eq:Nx1}
N_\x &\simeq \frac{\Gamma_\text{cap}}{K_\text{ann}} \, \tanh{\big( \sqrt{\Gamma_\text{cap} \, K_\text{ann}} ~ t_\odot  \big)}
\nonumber \\[3pt]
&\simeq \begin{cases}
\Gamma_\text{cap} \, t_\odot& (t_\odot \ll \tau)
\\[3pt]
\sqrt{\Gamma_\text{cap} / K_\text{ann}}& (t_\odot \gg \tau)
~,
\end{cases}
\end{align}
where $t_\odot \simeq 4.5 \ \text{Gyr}$ is the age of the Sun and $\tau \equiv (\Gamma_\text{cap} \, K_\text{ann})^{-1/2}$. From \Eqs{Nx1}{dNdt}, the annihilation rate grows as $N_\x^2 \, K_\text{ann} \sim \Gamma^2_\text{cap} \, K_\text{ann} \, t^2$ until $t \sim \tau$, after which equilibrium is reached between the rates of capture and annihilation, $N_\x^2 \, K_\text{ann} \simeq \Gamma_\text{cap}$.

\section{Millicharged Dark Matter Annihilation}
\label{sec:annihilation}

Millicharged particles can annihilate directly into pairs of charged Standard Model fermions through the exchange of an intermediate photon, $\x^+ \x^- \to \gamma^* \to f^+ f^-$. In the non-relativistic limit, the cross section for this processes is approximately given by 
\be
\label{eq:sigmaf}
\langle \sigma v\rangle_{\x^+ \x^- \rightarrow f^+ f^-} \simeq \frac{\pi \alpha^2 q_\x^2}{m_\x^2} \sum_f \, n_f \, q_f^2 
~,
\ee
where the sum is over all fermions with mass $m_f \ll m_\x$ and charge $q_f$, and $n_f$ is the number of colors of the fermion. Annihilations to electroweak bosons and neutrinos are also possible, but depend on the details of the full theory. For instance, if these interactions are generated by a kinetically-mixed dark photon, $\x^\pm$ will be millicharged under Standard Model hypercharge, such that it develops a small coupling to both the photon and the $Z$ boson. Hence, for $m_\x \gtrsim 10^2 \ \GeV$, annihilations to pairs of $W$ bosons and neutrinos are unsuppressed, further contributing to the signals discussed here. Since such processes are model-dependent, we have not included their contributions in our final estimate. 

In models that include a dark photon, $\Ap$, the dark matter can also annihilate to a pair of such particles. The cross section for this process is given by
\be
\label{eq:sigmaAp}
\langle \sigma v \rangle_{\x^+ \x^- \rightarrow \Ap \Ap} \simeq \frac{\pi \alpha^{\p \, 2}}{m_\x^2} 
~,
\ee
where $\mAp$ is the mass of the dark photon and $\alpha^\p = e^{\p \, 2} / 4 \pi$. If $\alpha^\p$ is sufficiently large, annihilations to dark photons could dominate over those to Standard Model final states, resulting in a suppression of the neutrino flux from the Sun as calculated in this study. In our analysis, we neglect this process, which is equivalent to taking $\alpha^\p \ll \alpha$.

\section{Bound States}
\label{sec:bound}

Negative-millicharged particles, $\x^-$, can efficiently bind to  nuclei, $N$, in the Sun, with a characteristic binding energy that is given by $E_{\x N} \simeq (q_\x Z \alpha)^2 \, \mu_{\x N} / 2$, where $Z$ is the atomic number of the nucleus and $\mu_{\x N}$ is the reduced mass of the system. Hence, for $m_\x \gtrsim \mu_{\x N} \sim 10 \ \GeV$ and $q_\x \gtrsim (m_e / \mu_{\x N})^{1/2} \sim 10^{-2}$, millicharged particles will be much more tightly-bound to nuclei than electrons are. Furthermore, for $Z \sim 10$ and similar values of the coupling $q_\x$, $E_{\x N}$ will exceed the temperature in the solar core, $T \sim 1 \ \keV$, in which case the formation of  $(\x^- N)$ bound states will be energetically favored.

Since $(\x^- N)$ has a net positive charge of $Q_{\x N} = Z - q_\x \simeq Z$, its formation suppresses the rate at which millicharged particles annihilate. In particular, such annihilations require that a $(\x^- N)$ and $\x^+$ overcome the repulsive Coulomb barrier, which is comparable to the binding energy, $Q_{\x N} \, q_\x \, \alpha / r_{\x N} \simeq E_{\x N}$, where $r_{\x N} \simeq (q_\x Z \alpha \, \, \mu_{\x N})^{-1}$ is the size of the bound state. As a result, if $E_{\x N} \gtrsim T$, then $(\x^- N)$ and $\x^+$ will not possess sufficient kinetic energy to overcome their electrical repulsion.\footnote{For sufficiently large $\alpha^\p$, the attractive dark Coulomb potential can overcome this repulsion. However, since this also suppresses the fraction of annihilations to Standard Model states (as discussed in \Sec{annihilation}), we will ignore this possibility.}

As the rates for the formation and disruption of $(\x^- N)$ bound states in the Sun are both very rapid~\cite{Pospelov:2020ktu}, chemical equilibrium will be reached and maintained between the bound, $(\x^- N)$, and unbound, $\x^-$, populations. From the condition of chemical equilibrium, $\mu_{\x^-} + \mu_N = \mu_{(\x^- N)}$, we can use the Maxwell-Boltzmann distributions of these species to relate the number density of $(\x^- N)$ bound states to the number densities of free nuclei and millicharged particles:
\be
\label{eq:ChemEq}
\frac{n_{\x^-} \, n_N}{n_{(\x^- N)}} =  \Big(\frac{\mu_{\x N} \, T}{2\pi}\Big)^{3/2} \,  e^{- E_{\x N} / T}
~.
\ee

Since annihilations between $\x^+$ and $(\x^- N)$ are exponentially suppressed by their repulsive Coulomb barrier, we incorporate the effect of bound state formation by rescaling the $\x^+ \x^-$ annihilation cross section by $\langle \sigma v \rangle \to \mathcal{R}_\text{BS} \, \langle \sigma v \rangle$, where
\be
\mathcal{R}_\text{BS} \equiv \frac{n_{\x^-} + e^{-E_{\x N} / T} \, n_{(\x^- N)}}{n_{\x^-} + n_{(\x^- N)}}
~.
\ee
Using \Eq{ChemEq}, this becomes
\begin{align}
\label{eq:RBS}
\mathcal{R}_\text{BS} &= \frac{F_N + 1}{F_N + \, e^{E_{\x N} / T}}
~,
\end{align}
where
\begin{align}
F_N &\equiv \Big(\frac{\mu_{\x N} \, T}{2\pi}\Big)^{3/2} \,  \frac{1}{n_N}
~.
\end{align}

From this calculation, we see that bound state formation exponentially suppresses the annihilation rate when $E_{\x N} \gtrsim T \, \ln{F_N}$. We incorporate this into our analysis by considering interactions with thorium (Th) nuclei, the highest-$Z$ element measured in the Sun, which constitutes $\sim 2\times 10^{-10}$ of the mass density in the photosphere~\cite{Asplund:2009fu}. We take this to be representative of the entire solar volume, with $n_N \simeq 7 \times 10^{11} \ \cm^{-3}$ independent of radius. We use this in \Eq{RBS} along with $T = 1 \ \keV$, which yields $\ln F_N \sim 50$ for $m_\x \gtrsim 1 \ \GeV$.

\begin{figure*}[t]
\centering
\includegraphics[width=0.495 \textwidth]{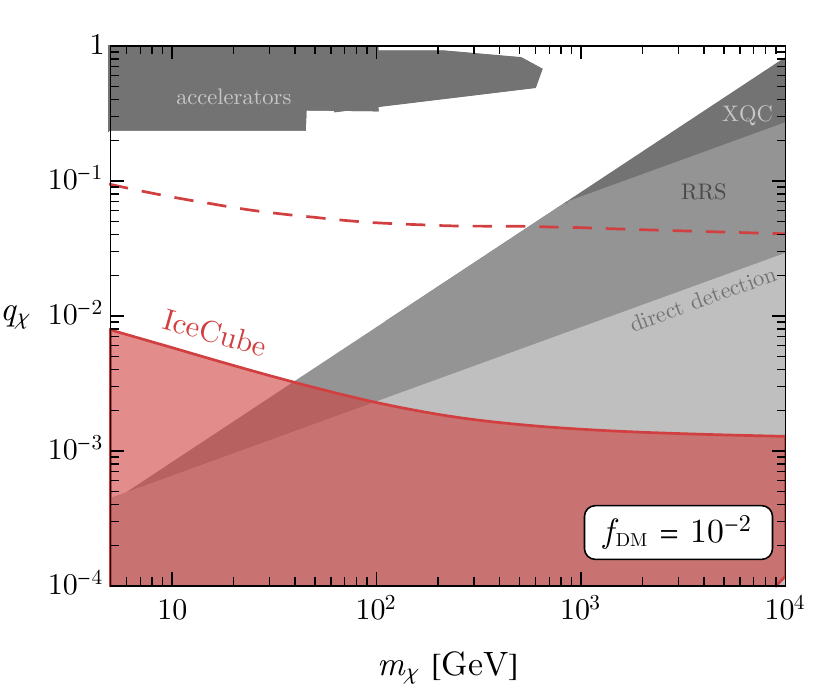}
\includegraphics[width=0.495 \textwidth]{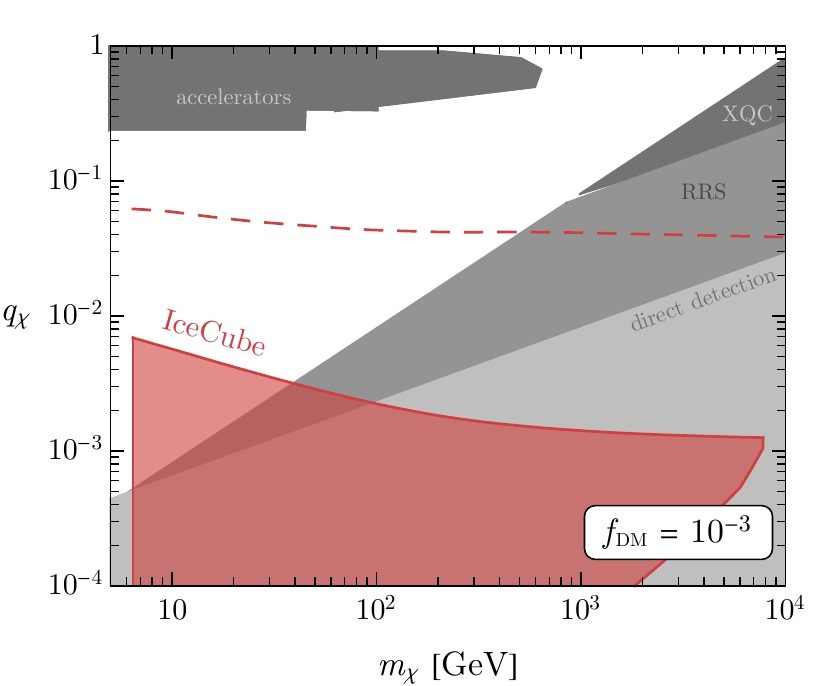}
\includegraphics[width=0.495 \textwidth]{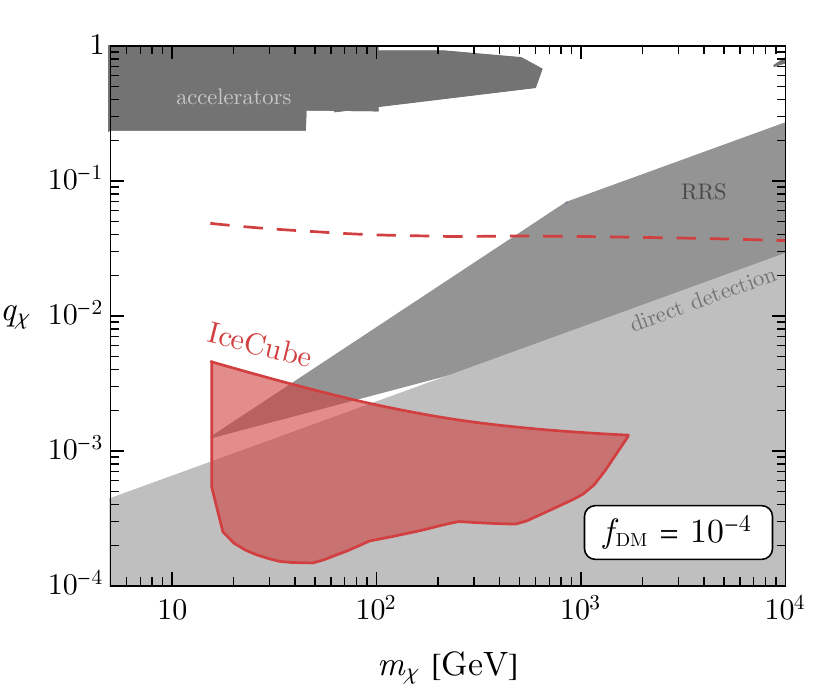}
\includegraphics[width=0.495 \textwidth]{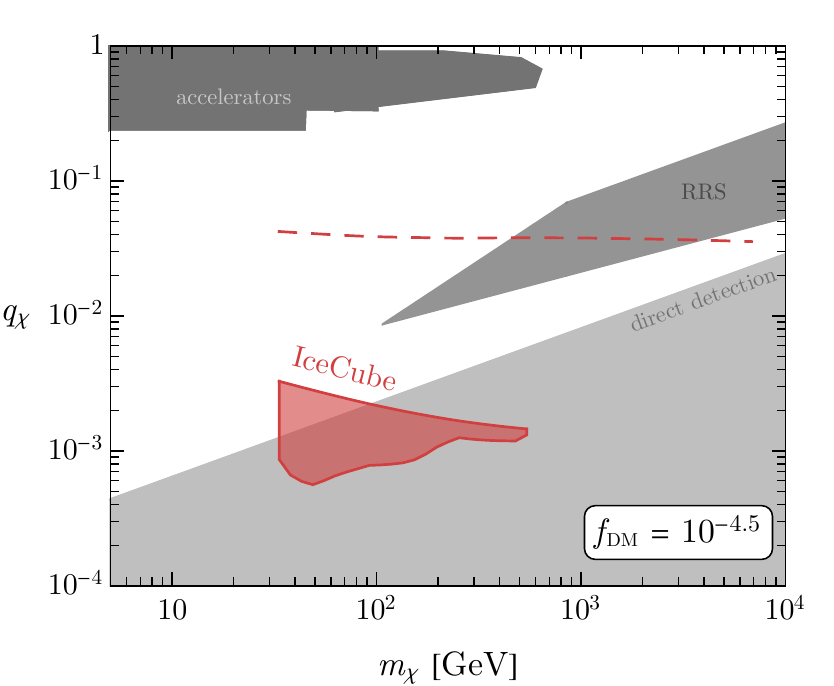}
\caption{In shaded red, regions of the millicharged dark matter parameter space that is excluded by our analysis, based on data from the IceCube Neutrino Observatory~\cite{IceCube:2016dgk,IceCube:2021xzo}, and for several choices of the fractional abundance, $f_\DM$. The dashed red lines denote the largest couplings that would have been excluded by our analysis if we had considered bound states with oxygen nuclei, rather than thorium (see \Sec{bound}). Also shown in gray are previous constraints from accelerator searches~\cite{Davidson:2000hf,CMS:2012xi,CMS:2013czn,CMS:2024eyx}, direct detection experiments~\cite{Emken:2019tni}, XQC~\cite{Erickcek:2007jv,Mahdawi:2018euy}, and RRS~\cite{Rich:1987st,Emken:2019tni}.  For RRS and XQC, we have extrapolated previous results to higher masses and rescaled to smaller fractional abundances; we have refrained from showing these limits when the gyroradius of millicharged dark matter in Earth's magnetic field is smaller than one tenth of an Earth radius.}
\label{fig:reach}
\end{figure*}
%

\section{Constraints From IceCube}
\label{sec:icecube}

The IceCube Neutrino Observatory consists of several thousand optical modules distributed throughout a cubic kilometer of Antarctic ice. In the energy range of interest, this instrument is primarily sensitive to muon tracks which result from the charged-current interactions of muon neutrinos in or near the detector's instrumented volume. Dark matter annihilating in the Sun could produce a potentially detectable flux of energetic neutrinos, allowing IceCube to place constraints on the particle nature of dark matter. 

The IceCube Collaboration has placed stringent limits on the flux of high-energy neutrinos that reaches us from the direction of the Sun~\cite{IceCube:2021xzo,IceCube:2016dgk,IceCube:2016yoy,IceCube:2012ugg,IceCube:2011aj,IceCube:2009iyf}. In this study, we recast the limits presented in Refs.~\cite{IceCube:2016dgk,IceCube:2021xzo}, which constrain the following quantity: 
\be
\label{eq:GammaAnn}
\Gamma_\text{ann} \equiv \frac{1}{2} \, N_\x^2 \, K_\text{ann}^\text{SM} = \frac{1}{4} \, \int_{V_\odot} \hspace{-0.1 cm} n_\x^2 \, \langle \sigma v \rangle_\text{SM}
\, dV~,
\ee
where $K_\text{ann}^\text{SM}$ is defined analogously to \Eq{Iann}, but in terms of the cross section for annihilations to a particular Standard Model final state, $\langle \sigma v \rangle_\text{SM}$. In this study, we focus on the case of $\tau^+ \tau^-$ final states, as these lead to strong bounds and do not depend on the complete model details of the theory (see the discussion in \Sec{annihilation}). 

Before proceeding, it is worthwhile to use the above results to estimate the signal rate in \Eq{GammaAnn}. This is simplest in the large coupling regime, in which case the capture and annihilation rate have reached steady state. In this case, Eqs.~\ref{eq:Acap}, \ref{eq:GammaCap1}, and \ref{eq:Nx1} can be used to evaluate this quantity: 
\be
\label{eq:GammaAnn2}
\Gamma_\text{ann} \simeq 5 \times 10^{29} \ \Hz \times f_\DM \, \bigg( \frac{\langle \sigma v \rangle_\text{SM}}{\langle \sigma v \rangle} \bigg) \,  \bigg( \frac{\GeV}{m_\x} \bigg)
\, .
\ee
Note that $\Gamma_\text{ann}$ is independent of $q_\x$, provided that the annihilations are dominated by Standard Model final states. As an example, for the annihilations of $m_\x = 100 \ \GeV$ millicharged particles to $\tau^+ \tau^-$, the search in Ref.~\cite{IceCube:2021xzo} places an upper limit of $\Gamma_\text{ann} \lesssim 4 \times 10^{21} \ \Hz$. Using the fact that $\langle \sigma v \rangle_{\tau^+ \tau^-} \sim 10^{-1} \times \langle \sigma v \rangle$, \Eq{GammaAnn2} then implies that IceCube is sensitive to fractional abundances as small as $f_\DM \sim \text{few} \times 10^{-5}$. 

This is confirmed in Fig.~\ref{fig:reach}, which presents our limits in the $q_\x - m_\x$ plane for various representative choices of the fractional abundance, $f_\DM$. Also shown are previous constraints from accelerator searches~\cite{Davidson:2000hf,CMS:2012xi,CMS:2013czn,CMS:2024eyx}, direct detection experiments~\cite{Emken:2019tni}, the rocket-based XQC experiment~\cite{Erickcek:2007jv,Mahdawi:2018euy}, and the balloon-based RRS experiment~\cite{Rich:1987st,Emken:2019tni}. For these latter two searches, we have extrapolated to higher masses and rescaled to smaller fractional abundances. We note that terrestrial dark matter searches cannot be simply extended to regions of parameter space in which millicharged particles strongly couple to Earth's magnetic field. In particular, we refrain from showing these limits when the gyroradius of $\x^\pm$ is smaller than $10^{-1} \, R_\oplus$, where $R_\oplus \simeq 6400 \ \km$ is the radius of Earth, since the millicharged particles will be significantly deflected before reaching terrestrial-based detectors in this case~\cite{Emken:2019tni}. We find that IceCube is sensitive to millicharged dark matter for masses in the range of $5 \ \GeV - 10 \ \TeV$ for fractional abundances as small as $f_\DM \sim 3 \times 10^{-5}$. This allows us to constrain unexplored parameter space for $m_\x \lesssim 10^2 \ \GeV$ and $q_\x \gtrsim 10^{-3}$, for which the couplings are too large to be probed by terrestrial-based experiments. 

For sufficiently large couplings, the formation of $(\x^- N)$ bound states suppresses the annihilation signals discussed here. This dictates the top of the regions probed by IceCube shown in \Fig{reach}. As discussed in \Sec{bound}, these limits are conservative, since we have incorporated thorium, the highest-$Z$ element measured in the photosphere, and have taken its photospheric density to be representative of its overall solar abundance. In contrast, the highest-$Z$ element for which we have an accurate understanding of its radial profile is oxygen, since it is actively produced in the Sun. The dashed lines shown in Fig.~\ref{fig:reach} represent the largest couplings that would be excluded by IceCube if only bound states with oxygen nuclei had been included in our analysis.  

Although not shown in \Fig{reach}, other bounds on millicharged dark matter subcomponents include those derived from ion traps~\cite{Budker:2021quh} (which are subject to the effect of Earth's magnetic field, as discussed above), spectral distortions of the CMB~\cite{Berlin:2022hmt} (which apply solely to models involving dark photons for parameter space that is complementary to that considered here), and white dwarf/neutron stars~\cite{Gould:1989gw,Fedderke:2019jur} (which are most relevant for masses larger than those considered in this work).

\section{Summary and Conclusions}
\label{sec:conclusion}

In this study, we have calculated the capture and annihilation rates of millicharged particles in the Sun, assuming that these particles make up a subdominant fraction of the total dark matter abundance. We have taken into account the effects of solar modulation, and the formation of $(\x^- N)$ bound states in the solar environment. From this annihilation rate and the resulting high-energy neutrino flux, we use previously published results from the IceCube Collaboration to constrain this class of models. We find that the IceCube data rules out significant regions of previously unexplored parameter space, featuring $m_\x \sim (5-100) \ \GeV$, $q_\x \sim 10^{-3}-10^{-2}$, and fractional abundances as small as $f_\DM \sim  3 \times 10^{-5}$.


\section*{Acknowledgements}
Fermilab is operated by the Fermi Research Alliance, LLC under Contract DE-AC02-07CH11359 with the U.S. Department of Energy. This material is based upon work supported by the U.S. Department of Energy, Office of Science, National Quantum Information Science Research Centers, Superconducting Quantum Materials and Systems Center (SQMS) under contract number DE-AC02-07CH11359. This work was completed in part at the Perimeter Institute. Research at Perimeter Institute is supported in part by the Government of Canada through the Department of Innovation, Science and Economic Development Canada and by the Province of Ontario through the Ministry of Colleges and Universities.


\bibliography{biblio}


\end{document}